\documentstyle[aps,preprint,epsf]{revtex}

\ifx\epsffile\undefined
\def\insertfig#1{}
\else
\def\insertfig#1{{\baselineskip=4pt
\centerline{\epsfxsize=\hsize\epsffile{#1}}}}\fi

\begin{document}

{\tighten
\preprint{\vbox{
\hbox{UCSD/PTH 96--28}
\hbox{hep-ph/9612211}
}}
\title{The Rare Top Decays $t \rightarrow b W^+ Z$ and 
$t \rightarrow c W^+ W^-$}
\author{Elizabeth Jenkins\footnotemark\footnotemark}
\address{Department of Physics, University of California at
San Diego, La Jolla, CA 92093}
\bigskip
\date{November 1996}
\maketitle
\widetext
\begin{abstract}
The large value of the top quark mass implies that the rare top decays
$t \rightarrow b W^+ Z$, $s W^+ Z$ and $d W^+ Z$,
and $t \rightarrow c W^+ W^-$ and $u W^+ W^-$,
are kinematically allowed so long as $m_t \ge m_W + m_Z + m_{d_i}
\approx 171.5~{\rm GeV} + m_{d_i}$ or 
$m_t \ge 2m_W + m_{u,c} \approx 160.6~{\rm GeV} + m_{u,c}$, respectively.  
The partial decay widths for these decay
modes are calculated in the standard model.  The partial widths depend
sensitively on the precise value of the top quark mass.  The branching
ratio for $t\rightarrow b W^+ Z$ is as much as $1 \times 10^{-5}$ for
$m_t = 200$~GeV, and could be observable at LHC.  The rare decay modes 
$t \rightarrow c W^+ W^-$ and 
$u W^+ W^-$ are highly GIM-suppressed, and thus provide  
a means for testing the GIM mechanism for three generations of quarks
in the $u$, $c$, $t$ sector.  
\end{abstract}
}

\footnotetext{${}^*$Alfred P. Sloan Fellow.}
\footnotetext{${}^\dagger$National Science Foundation Young Investigator.}

Now that the top quark mass is known to be quite large, it is possible to
examine the question of which rare decay modes of the top are kinematically
allowed processes.  The current CDF and D0 average value of the top quark
mass $m_t = 175 \pm 8$~GeV\cite{cd,cdfd0} implies that the decays 
$t \rightarrow b W^+ Z$, $s W^+ Z$ and $d W^+ Z$,
are allowed decay modes of the top so long as $m_t \ge m_W + m_Z + m_{d_i}
\approx 171.5~{\rm GeV} + m_{d_i}$.  The rare decays
$t \rightarrow c W^+ W^-$ and $u W^+ W^-$ also are allowed if
$m_t \rightarrow 2 m_W + m_{c,u} \approx 160.6~{\rm GeV} + m_{c,u}$.
For the present central value of the
measured top quark mass, all of these processes are occurring at or near 
threshold, and are highly phase space suppressed.  The decays  
$t \rightarrow c  W^+ W^-$ and $u W^+ W^-$
are also highly GIM-suppressed, and thus are
not likely to be seen at standard model rates.  The decays $t \rightarrow d_i
W^+ Z$, however, are not GIM-suppressed and are potentially observable at
the LHC.  The partial decay widths for these rare
decay modes rapidly increase for larger values of the top quark mass, 
and thus are very sensitive to the precise value of the top quark mass.
Since the decay widths are proportional to $|V_{td_i}|^2$, $i=1,2,3$, the
rare decay $t \rightarrow b W^+ Z$ (with $|V_{tb}|^2 \approx 1$)
will dominate unless the value of the
top quark is below or nearly at threshold for this process.  

We begin with
the calculation of the partial decay width $\Gamma(t \rightarrow b W^+ Z)$ in
the standard model.  The branching ratio for this decay process 
has been computed previously by
Decker, Nowakowski and Pilaftsis\cite{decker} and 
Mahlon and Parke\cite{mahlon}.\footnote{The decay process $Q \rightarrow qWZ$ also was
considered in Ref.~\cite{rizzo} for very heavy fourth generation quarks and
exotics with mass $\ge 240$~GeV.}  The authors of
Ref.~\cite{mahlon} included the finite
widths of the $W$ and $Z$ in their calculation, and found a significant 
enhancement in the decay width near threshold due to
finite width effects.   
There is some disagreement in the
numerical results of Ref.~\cite{decker} and \cite{mahlon}.  The numerical
results presented here are basically consistent with the published results of
Ref.~\cite{mahlon} in the narrow width approximation.  There is
some numerical difference with Ref.~\cite{mahlon}
which probably stems from the inclusion of
finite width effects in that calculation.
In addition, explicit analytic formulae
for the squared amplitude of $t \rightarrow b W^+ Z$
are presented in this work.  These formulae do not
appear elsewhere in the literature, and are useful for 
more detailed studies of the decay mode.
Finally, 
the decay widths for the other rare decay modes $t \rightarrow c W^+ W^-$ and
$u W^+ W^-$ also
are computed.  A search for these decay modes directly tests CKM
unitarity in the $u$-quark sector.

\section{$t \rightarrow b W^+ Z$}

The rare decay $t \rightarrow b W^+ Z$ proceeds via the three tree-level
graphs drawn in Fig.~1.  The amplitudes for these Feynman diagrams are   
\begin{eqnarray}
{\cal A}_1 &=& V_{tb} \left({{ig} \over \sqrt{2}}\right)
\left({{ig} \over {\cos{\theta_W}}}\right) \epsilon_W^\mu \epsilon_Z^\nu \ 
\bar u(p_b) \left[ \gamma^\mu P_L \left({i \over {{\rlap{$k_1$}/} - m_t}}\right)
\left\{ g_{t_L} \gamma^\nu P_L + g_{t_R} \gamma^\nu P_R \right\} \right] 
u(p_t), \\
{\cal A}_2 &=& V_{tb} \left({{ig} \over \sqrt{2}}\right)
\left({{ig} \over {\cos{\theta_W}}}\right) \epsilon_W^\mu \epsilon_Z^\nu \ 
\bar u(p_b) \left[ \left\{ g_{b_L} \gamma^\nu P_L 
+ g_{b_R} \gamma^\nu P_R \right\}
\left({i \over {{\rlap{$k_2$}/} - m_b}}\right)
\gamma^\mu P_L \right] u(p_t), \\
{\cal A}_3 &=& V_{tb} \left({{ig} \over \sqrt{2}}\right)
\left( {ig} \cos{\theta_W} \right) \epsilon_W^\mu \epsilon_Z^\nu \ 
\left({ {-i} \over {k_3^2 - m_W^2} } \right)
\left( g^{\lambda \rho} - {{k_3^\lambda k_3^\rho} \over m_W^2 } \right) 
\bar u(p_b) \gamma^\lambda P_L u(p_t) \times\\ 
&&\qquad\qquad
\left[ -g^{\mu \rho} \left( k_3 + p_W \right)^\nu + g^{\nu \rho}
\left( p_Z + k_3 \right)^\mu + g^{\mu \nu} \left( - p_Z + p_W \right)^\rho
\right],\nonumber
\end{eqnarray}
where the four momenta $k_1$, $k_2$ and $k_3$ are given by
\begin{eqnarray}
k_1 &=& p_t - p_Z = p_b + p_W, \nonumber\\
k_2 &=& p_t - p_W = p_b + p_Z, \\
k_3 &=& p_t - p_b = p_W + p_Z, \nonumber
\end{eqnarray}
and the couplings of the $Z$ boson to the left- and right-handed top and bottom
quarks are
\begin{eqnarray}
g_{t_L} &=& \left( \frac 1 2 - \frac 2 3 \sin^2{\theta_W} \right), \nonumber\\
g_{t_R} &=& \left( - \frac 2 3 \sin^2{\theta_W} \right), \nonumber\\
g_{b_L} &=& \left( - \frac 1 2 + \frac 1 3 \sin^2{\theta_W} \right), \\
g_{b_R} &=& \left( \frac 1 3 \sin^2{\theta_W} \right). \nonumber
\end{eqnarray}
In the above amplitudes, $P_{L,R}$ stand for the left- and right-handed 
projectors $P_{L,R} = ( 1 \mp \gamma_5 )/2$, and $\theta_W$ is the
weak mixing angle.  The amplitude ${\cal A}_3$ depends on the triple gauge
vertex $W^+ W^- Z$.  This amplitude has been written in unitary gauge,
where there is a contribution to the $W$ gauge boson propagator proportional
to $k_3^\lambda k_3^\rho/m_W^2$.  The amplitude also can be written in
t'Hooft-Feynman gauge ($\xi = 1$), where this contribution is replaced by
the exchange of the would-be Goldstone boson of the $W$.  

The total amplitude is given by ${\cal A} ={\cal A}_1 +{\cal A}_2 +{\cal A}_3$,
and the amplitude squared is
\begin{equation}
| {\cal A} |^2 = |{\cal A}_1|^2+|{\cal A}_2|^2+|{\cal A}_3|^2
+2 {\cal A}_1 {\cal A}_2^* +2 {\cal A}_1 {\cal A}_3^*
+ 2 {\cal A}_2 {\cal A}_3^*,
\end{equation}
where the identities 
\begin{eqnarray}
{\cal A}_1 {\cal A}_2^* &=& {\cal A}_2 {\cal A}_1^*, \nonumber\\
{\cal A}_1 {\cal A}_3^* &=& {\cal A}_3 {\cal A}_1^*, \\
{\cal A}_2 {\cal A}_3^* &=& {\cal A}_3 {\cal A}_2^*, \nonumber
\end{eqnarray}
have been used.

The square amplitude $|{\cal A}_1|^2$ is  
\begin{eqnarray}
|{\cal A}_1|^2 &=& |V_{tb}|^2 \left( { g^4 \over {2 \cos^2\theta_W} } \right)
\left( { 1 \over {k_1^2 - m_t^2} } \right)^2 \times \nonumber\\
&&\Bigg( 4 g_{t_L}^2 \bigg\{ 
\left[ \left( k_1 \cdot p_t \right) \left( k_1 \cdot p_b \right)
- \frac 1 2 k_1^2 \left( p_t \cdot p_b \right) \right]\nonumber\\
&&\qquad+{2 \over m_Z^2} \left( p_t \cdot p_Z \right) 
\left[ \left( k_1 \cdot p_b \right) \left( k_1 \cdot p_Z \right)
- \frac 1 2 k_1^2  \left( p_b \cdot p_Z \right) \right] \nonumber\\
&&\qquad+{2 \over m_W^2} \left( p_b \cdot p_W \right) 
\left[ \left( k_1 \cdot p_t \right) \left( k_1 \cdot p_W \right)
- \frac 1 2 k_1^2  \left( p_t \cdot p_W \right) \right] \nonumber\\
&&\qquad+ { {4} \over {m_W^2 m_Z^2}}
\left(p_t \cdot p_Z \right) \left(p_b \cdot p_W \right) 
\left[ \left( k_1 \cdot p_Z \right) \left( k_1 \cdot p_W \right)
-\frac 1 2 k_1^2 \left(p_W \cdot p_Z \right) \right] \bigg\} \nonumber\\
&&-12 m_t^2 g_{t_L} g_{t_R} \left\{ \left( k_1 \cdot p_b \right) 
+{2 \over m_W^2}
\left( k_1 \cdot p_W\right)\left(p_b \cdot p_W\right) \right\}\nonumber\\
&&+2 m_t^2 g_{t_R}^2 \left\{ \left(p_t \cdot p_b \right)
+ {2 \over m_Z^2} \left( p_t \cdot p_Z \right) \left(p_b \cdot p_Z \right)
+ {2 \over m_W^2} \left( p_t \cdot p_W \right) \left(p_b \cdot p_W \right)
\right.\nonumber\\
&&\quad\qquad+\left. 
{4 \over {m_W^2 m_Z^2}} \left(p_t \cdot p_Z\right) 
\left(p_b \cdot p_W \right) \left(p_W \cdot p_Z \right) \right\}\Bigg)
\end{eqnarray}
The square amplitude $|{\cal A}_2|^2$ is related to $|{\cal A}_1|^2$
by the interchanges $g_{t_{L,R}} \leftrightarrow g_{b_{L,R}}$,
$m_t \leftrightarrow m_b$, $p_t \leftrightarrow p_b$ and 
$k_1 \leftrightarrow k_2$,
\begin{eqnarray}
|{\cal A}_2|^2 &=& |V_{tb}|^2 \left( { g^4 \over {2 \cos^2\theta_W} } \right)
\left( { 1 \over {k_2^2 - m_b^2} } \right)^2 \times \nonumber\\
&&\Bigg( 4 g_{b_L}^2 \bigg\{ 
\left[\left( k_2 \cdot p_b \right) \left( k_2 \cdot p_t \right)
- \frac 1 2 k_2^2 \left( p_t \cdot p_b \right) \right]\nonumber\\
&&\qquad+{2 \over m_Z^2} \left( p_b \cdot p_Z \right) 
\left[ \left( k_2 \cdot p_t \right) \left( k_2 \cdot p_Z \right)
- \frac 1 2 k_2^2  \left( p_t \cdot p_Z \right) \right] \nonumber\\
&&\qquad+{2 \over m_W^2} \left( p_t \cdot p_W \right) 
\left[ \left( k_2 \cdot p_b \right) \left( k_2 \cdot p_W \right)
- \frac 1 2 k_2^2  \left( p_b \cdot p_W \right) \right] \nonumber\\
&&\qquad+ { {4} \over {m_W^2 m_Z^2}}
\left(p_b \cdot p_Z \right) \left(p_t \cdot p_W \right) 
\left[ \left( k_2 \cdot p_Z \right) \left( k_2 \cdot p_W \right)
-\frac 1 2 k_2^2 \left(p_W \cdot p_Z \right) \right] \bigg\} \nonumber\\
&&-12 m_b^2 g_{b_L} g_{b_R} \left\{ \left( k_2 \cdot p_t \right) 
+{2 \over m_W^2}
\left( k_2 \cdot p_W\right)\left(p_t \cdot p_W\right) \right\}\nonumber\\
&&+2 m_b^2 g_{b_R}^2 \left\{ \left(p_t \cdot p_b \right)
+ {2 \over m_Z^2} \left( p_b \cdot p_Z \right) \left(p_t \cdot p_Z \right)
+ {2 \over m_W^2} \left( p_b \cdot p_W \right) \left(p_t \cdot p_W \right)
\right.\nonumber\\
&&\quad\qquad+\left. 
{4 \over {m_W^2 m_Z^2}} \left(p_b \cdot p_Z\right) 
\left(p_t \cdot p_W \right) \left(p_W \cdot p_Z \right) \right\}\Bigg)\ .
\end{eqnarray}

The square amplitude $|{\cal A}_3|^2$ is 
\begin{eqnarray}
|{\cal A}_3|^2 &&= |V_{tb}|^2 \left( { {g^4 \cos^2\theta_W} \over {2} } \right)
\left( { 1 \over {k_3^2 - m_W^2} } \right)^2 \times \nonumber\\
&&\Bigg( 4 \left( p_t \cdot p_b \right) \bigg\{ - 3 \left(m_W^2 + m_Z^2
\right) + 2 \gamma \left( m_W^2 - m_Z^2 \right) - \gamma^2 \left( m_W^2
+ m_Z^2 \right) 
+ 2 \left( 1 - \gamma^2 \right) \left( p_W \cdot p_Z \right) \nonumber\\
&&\qquad\qquad+ \left[ {3 \over 2} \left( {1 \over m_W^2} 
+ {1 \over m_Z^2} \right)
+ \gamma \left( {1 \over m_Z^2} - {1 \over m_W^2} \right) 
-{1 \over 2} \gamma^2 \left( {1 \over m_W^2} + {1 \over m_Z^2} \right) \right]
\left( p_W \cdot p_Z \right)^2 \nonumber\\
&&\qquad\qquad+ {1 \over {m_W^2 m_Z^2}}
\left( 1 - \gamma^2 \right)
\left( p_W \cdot p_Z \right)^3 
\bigg\} \\
&&+4\left(p_t \cdot p_W\right)\left(p_b \cdot p_W \right) \left\{
-2 -4 \gamma +2 \gamma^2 + {4 \over m_W^2}\left( 1 + \gamma \right) 
\left(p_W \cdot p_Z \right)
+ {1 \over {m_W^2 m_Z^2}} \left( 1 + 2 \gamma + \gamma^2 \right)
\left(p_W \cdot p_Z \right)^2 \right\}
\nonumber\\
&&+4\left(p_t \cdot p_Z\right)\left(p_b \cdot p_Z \right) \left\{
-2 + 4 \gamma + 2 \gamma^2 + {{4} \over m_Z^2} \left( 1 - \gamma \right) 
\left( p_W \cdot p_Z \right) 
+ {{1} \over {m_W^2 m_Z^2} }\left( 1 - 2 \gamma + \gamma^2 \right)
\left( p_W \cdot p_Z \right)^2 \right\}
\nonumber\\
&&+4\left[ \left(p_t \cdot p_W \right) \left( p_b \cdot p_Z \right) +
\left( p_b \cdot p_W \right) \left(p_t \cdot p_Z \right) \right]
\bigg\{-6 + 2 \gamma^2 \nonumber\\
&&\qquad+ \left[2 \left( {1 \over m_Z^2} + {1 \over m_W^2} \right)
-2 \gamma \left( {1 \over m_Z^2} - {1 \over m_W^2} \right) \right]
\left( p_W \cdot p_Z \right)
-{1 \over {m_W^2 m_Z^2}}\left( 1 - \gamma^2 \right) 
\left( p_W \cdot p_Z \right)^2 \bigg\}
\Bigg)\ ,\nonumber
\end{eqnarray}
where $\gamma = \left( 1 - m_Z^2/m_W^2 \right)$ in unitary gauge.  In
t'Hooft-Feynman gauge, one obtains the same expression with $\gamma =
\sin^2\theta_W$ and $\cos^2\theta_W$ replaced by $(m_Z/m_W)^2$.
The terms in $|{\cal A}_3|^2$ proportional to $\gamma$  
are antisymmetric under $p_t \leftrightarrow p_b$,
$p_W \leftrightarrow p_Z$ and $m_W \leftrightarrow m_Z$ while the
terms which are independent of $\gamma$ or proportional to $\gamma^2$
are invariant under this interchange.

The interference term ${\cal A}_1 {\cal A}_2^*$ is
\begin{eqnarray}
{\cal A}_1 {\cal A}_2^* &=& |V_{tb}|^2 
\left( { g^4 \over {2 \cos^2\theta_W} } \right)
\left( { 1 \over {k_1^2 - m_t^2} } \right) 
\left( { 1 \over {k_2^2 - m_b^2} } \right) \times \nonumber\\
&&\Bigg( g_{t_L} g_{b_L} \bigg\{
-6 \left( k_1 \cdot k_2 \right)\left(p_t \cdot p_b\right)
-2 \left( k_1 \cdot p_t \right)\left(k_2 \cdot p_b \right)
-2 \left( k_1 \cdot p_b \right)\left(k_2 \cdot p_t \right) \nonumber\\
&&-{4 \over m_Z^2} \Bigg[ \left(k_1 \cdot k_2 \right) 
\left(p_b \cdot p_Z \right)
\left(p_t \cdot p_Z \right) 
+ \left( k_1 \cdot p_Z \right)\left(k_2 \cdot p_Z \right) 
\left(p_t \cdot p_b \right)
+\left( k_2 \cdot p_Z \right) \left(k_1 \cdot p_t \right)\left(p_b \cdot p_Z
\right)\nonumber\\ 
&&+ \left(k_1 \cdot p_Z \right) \left(p_t \cdot p_Z \right)\left(k_2
\cdot p_b \right)-2 \left(p_t \cdot p_Z \right)\left(k_1 \cdot p_b \right)
\left(k_2 \cdot p_Z \right) -2 \left(k_1 \cdot p_Z \right)\left(p_b \cdot p_Z
\right)\left(k_2 \cdot p_t \right) \Bigg]\nonumber\\
&&-{4 \over m_W^2} \Bigg[ \left(k_1 \cdot k_2 \right) \left(p_b \cdot p_W \right)
\left(p_t \cdot p_W \right) 
+ \left( k_1 \cdot p_W \right)\left(k_2 \cdot p_W \right) 
\left(p_t \cdot p_b \right)
+\left( k_2 \cdot p_W \right) \left(k_1 \cdot p_b \right)\left(p_t \cdot p_W
\right) \nonumber\\
&&+ \left(k_1 \cdot p_W \right) \left(p_b \cdot p_W \right)\left(k_2
\cdot p_t \right)-2 \left(p_b \cdot p_W \right)\left(k_1 \cdot p_t \right)
\left(k_2 \cdot p_W \right) -2 \left(k_1 \cdot p_W \right)\left(p_t \cdot p_W
\right)\left(k_2 \cdot p_b \right) \Bigg]\nonumber\\
&&+{4 \over {m_W^2 m_Z^2}}\Bigg[\left(k_1 \cdot k_2 \right)\left(p_W \cdot p_Z
\right)\left\{
-\left(p_t \cdot p_b\right)\left(p_W \cdot p_Z \right) +\left(p_b \cdot p_Z
\right)\left(p_t \cdot p_W \right)+\left(p_b \cdot p_W \right)\left(p_t \cdot
p_Z \right) \right\} \nonumber\\
&&+ \left( p_W \cdot p_Z \right)^2 \left\{ \left(k_1 \cdot p_t \right)
\left(k_2 \cdot p_b \right) + \left(k_1 \cdot p_b \right) \left(k_2 \cdot p_t
\right) \right\}\nonumber\\
&&+\left(p_t \cdot p_b \right)\left(p_W \cdot p_Z\right) \left\{\left(k_1 \cdot
p_Z \right)\left(k_2 \cdot p_W \right) + \left(k_1 \cdot p_W \right)\left(k_2
\cdot p_Z \right) \right\}\nonumber\\
&&+2\left(k_1 \cdot p_Z \right)\left(k_2 \cdot p_Z \right)\left(p_b \cdot
p_W\right)\left(p_t \cdot p_W \right) 
+2\left(k_1 \cdot p_W \right)\left(k_2 \cdot p_W \right)\left(p_b \cdot
p_Z\right)\left(p_t \cdot p_Z \right)\nonumber\\ 
&&-\left(k_1 \cdot p_b \right)\left(k_2 \cdot p_Z \right)\left(p_t \cdot p_W
\right) \left(p_W \cdot p_Z \right) - \left(k_2 \cdot p_t \right)\left(k_1
\cdot p_Z \right)\left( p_b \cdot p_W \right)\left(p_W \cdot p_Z \right)
\nonumber\\
&&-\left(k_1 \cdot p_t \right)\left(k_2 \cdot p_Z\right)\left(p_b \cdot p_W
\right)\left(p_W \cdot p_Z \right) 
-\left(k_2 \cdot p_b \right)\left(k_1 \cdot p_Z\right)\left(p_t \cdot p_W
\right)\left(p_W \cdot p_Z \right)\nonumber\\ 
&&-\left(k_1 \cdot p_t \right)\left(k_2 \cdot p_W\right)\left(p_b \cdot p_Z
\right)\left(p_W \cdot p_Z \right) 
-\left(k_2 \cdot p_b \right)\left(k_1 \cdot p_W\right)\left(p_t \cdot p_Z
\right)\left(p_W \cdot p_Z \right)\nonumber\\ 
&&-\left(k_1 \cdot p_b \right)\left(k_2 \cdot p_W\right)\left(p_W \cdot p_Z
\right)\left(p_t \cdot p_Z \right) 
-\left(k_2 \cdot p_t \right)\left(k_1 \cdot p_W\right)\left(p_b \cdot p_Z
\right)\left(p_W \cdot p_Z \right) \Bigg]\bigg\}\nonumber\\
&&+2 m_b^2 g_{t_L} g_{b_R} \bigg\{ 3 \left( k_1 \cdot p_t \right) + 
{2 \over m_Z^2} \left( k_1 \cdot p_Z \right) \left( p_t \cdot p_Z \right)
-{4 \over m_W^2} \left( p_t \cdot p_W \right) \left( k_1 \cdot p_W \right)
\nonumber\\
&&+ {2 \over {m_W^2 m_Z^2 }}\left( p_W \cdot p_Z \right) 
\left[ \left(k_1 \cdot p_W \right) \left( p_t \cdot p_Z \right) + \left( p_t
\cdot p_W \right) \left( k_1 \cdot p_Z \right) 
-\left( p_W \cdot p_Z \right) \left(k_1 \cdot p_t \right) \right] \bigg\}
\nonumber\\
&&+2 m_t^2 g_{t_R} g_{b_L} \bigg\{ 3 \left( k_2 \cdot p_b \right) + 
{2 \over m_Z^2} \left( k_2 \cdot p_Z \right) \left( p_b \cdot p_Z \right)
-{4 \over m_W^2} \left( p_b \cdot p_W \right) \left( k_2 \cdot p_W \right)
\nonumber\\
&&+ {2 \over {m_W^2 m_Z^2 }}\left( p_W \cdot p_Z \right) 
\Big[ \left(k_2 \cdot p_W \right) \left( p_b \cdot p_Z \right) + \left( p_b
\cdot p_W \right) \left( k_2 \cdot p_Z \right) 
-\left( p_W \cdot p_Z \right) \left(k_2 \cdot p_b \right) \Big] \bigg\}
\nonumber\\
&&+2 m_t^2 m_b^2 g_{t_R} g_{b_R} \left\{ -5 
+ {2 \over {m_W^2 m_Z^2}} \left( p_W \cdot p_Z \right)^2 \right\}
\Bigg)
\end{eqnarray}
${\cal A}_1 {\cal A}_2^*$ is invariant under the simultaneous interchanges
$g_{t_{L,R}} \leftrightarrow g_{b_{L,R}}$, $m_t \leftrightarrow m_b$, 
$p_t \leftrightarrow p_b$ and $k_1 \leftrightarrow k_2$.

The two interference terms ${\cal A}_1 {\cal A}_3^*$ and
${\cal A}_2 {\cal A}_3^*$ are
\begin{eqnarray}
{\cal A}_1 {\cal A}_3^* &=& |V_{tb}|^2 
\left( { {g^4} \over {2} } \right)
\left( { 1 \over {k_1^2 - m_t^2} } \right) 
\left( { 1 \over {k_3^2 - m_W^2} } \right) \times \nonumber\\
&&\Bigg( g_{t_L} 
\bigg\{ -4 \left[ \left( 2 -\gamma \right)
\left(k_1 \cdot p_W \right)\left(p_t \cdot p_b \right) 
+ \left(k_1 \cdot p_b \right)\left(p_t \cdot p_W \right) -3 \left(k_1
\cdot p_t \right) \left(p_b \cdot p_W \right)\right]\nonumber\\
&&\qquad+4 \left[ \left( 2 + \gamma \right)
\left(k_1 \cdot p_Z \right)\left(p_t \cdot p_b
\right) 
+ \left(k_1 \cdot p_t \right)\left(p_b \cdot p_Z \right) -3 \left(k_1
\cdot p_b \right) \left(p_t \cdot p_Z \right)\right]\nonumber\\
&&-{4 \over m_W^2}\left( 1 + \gamma \right)
\left(k_1 \cdot p_W\right)
\left[ 2 \left(p_t \cdot p_W \right) \left(p_b \cdot p_W \right) 
+\left(p_t \cdot p_W \right)\left(p_b \cdot p_Z \right) 
\right. \nonumber\\ 
&&\left.\qquad\qquad\qquad\qquad\qquad\qquad
-\left(p_W \cdot p_Z \right)\left(p_t \cdot p_b \right)
+\left(p_b \cdot p_W \right)\left(p_t \cdot p_Z \right) \right]\nonumber\\
&&+{4 \over m_Z^2}\left( 1 - \gamma \right)
\left(k_1 \cdot p_Z\right)\left[ 2 \left(p_t \cdot p_Z
\right) \left(p_b \cdot p_Z \right) + \left(p_t \cdot p_Z \right)\left(p_b
\cdot p_W \right) \right. \nonumber\\ 
&&\left.\qquad\qquad\qquad\qquad\qquad\qquad 
- \left(p_W \cdot p_Z \right)\left(p_t \cdot p_b \right)
+ \left(p_b \cdot p_Z \right)\left(p_t \cdot p_W \right) \right]\nonumber\\
&&-{{2} \over m_W^2}\left(3 + \gamma \right)
\left(p_W \cdot p_Z \right)\left[
\left(k_1 \cdot p_W\right)\left(p_t \cdot p_b \right) - \left(p_t \cdot p_W
\right)\left(k_1 \cdot p_b \right) + \left( p_b \cdot p_W \right) \left(k_1
\cdot p_t \right) \right] \nonumber\\
&&+{{2} \over m_Z^2}\left(3 - \gamma \right)
\left(p_W \cdot p_Z \right)\left[
\left(k_1 \cdot p_Z\right)\left(p_t \cdot p_b \right) - \left(p_b \cdot p_Z
\right)\left(k_1 \cdot p_t \right) + \left( p_t \cdot p_Z \right) \left(k_1
\cdot p_b \right) \right] \nonumber\\
&&+{4 \over {m_W^2 m_Z^2}}
\left( 1 + \gamma \right)
\left(p_W \cdot p_Z\right)\left(p_b \cdot p_W \right)
\left[\left(k_1 \cdot p_W \right)\left(p_t \cdot p_Z \right) - \left(p_W \cdot
p_Z \right) \left(k_1 \cdot p_t \right)+ \left(p_t \cdot p_W \right) \left(k_1
\cdot p_Z \right) \right]\nonumber\\
&&-{4 \over {m_W^2 m_Z^2}}
\left( 1 - \gamma \right)
\left(p_W \cdot p_Z\right)\left(p_t \cdot p_Z \right)
\left[\left(k_1 \cdot p_W \right)\left(p_b \cdot p_Z \right) - \left(p_W \cdot
p_Z \right) \left(k_1 \cdot p_b \right)+ \left(p_b \cdot p_W \right) \left(k_1
\cdot p_Z \right) \right]\bigg\}\nonumber\\
&&+4 m_t^2 g_{t_R} 
\left\{ \left(-4 + \gamma \right)
\left(p_b \cdot p_W \right) 
+ \left(-2 + \gamma \right)
\left(p_b \cdot p_Z \right)
+{{1} \over {m_W^2}}\left( {7 \over 2} + {1 \over 2} \gamma \right)
\left(p_W \cdot p_Z \right)\left(p_b \cdot p_W \right) \right.\nonumber\\ 
&&\left.\qquad\qquad
+{1 \over m_Z^2} \left( {3 \over 2} -{1 \over 2} \gamma \right)
\left(p_W \cdot p_Z \right)\left(p_b \cdot p_Z \right)
+{1 \over {m_W^2 m_Z^2}}\left(1 + \gamma \right)
\left(p_W \cdot p_Z \right)^2 \left(p_b \cdot p_W \right)\right\}
\end{eqnarray}
and
\begin{eqnarray}
{\cal A}_2 {\cal A}_3^* &=& |V_{tb}|^2 
\left( { {g^4} \over {2} } \right)
\left( { 1 \over {k_2^2 - m_b^2} } \right) 
\left( { 1 \over {k_3^2 - m_W^2} } \right) \times \nonumber\\
&&\Bigg( g_{b_L} 
\bigg\{ -4 \left[ \left(2 - \gamma \right)
\left(k_2 \cdot p_W \right)\left(p_t \cdot p_b
\right) + \left(k_2 \cdot p_t \right)\left(p_b \cdot p_W \right) -3 \left(k_2
\cdot p_b \right) \left(p_t \cdot p_W \right)\right]\nonumber\\
&&\qquad +4 \left[ \left(2 + \gamma \right)
\left(k_2 \cdot p_Z \right)\left(p_t \cdot p_b
\right) + \left(k_2 \cdot p_b \right)\left(p_t \cdot p_Z \right) -3 \left(k_2
\cdot p_t \right) \left(p_b \cdot p_Z \right)\right]\nonumber\\
&&-{4 \over m_W^2}\left(1 + \gamma \right)
\left(k_2 \cdot p_W\right)\left[ 2 \left(p_b \cdot p_W
\right) \left(p_t \cdot p_W \right) + \left(p_b \cdot p_W \right)\left(p_t
\cdot p_Z \right) 
\right. \nonumber\\ 
&&\left.\qquad\qquad\qquad\qquad\qquad\qquad
- \left(p_W \cdot p_Z \right)\left(p_t \cdot p_b \right)
+ \left(p_t \cdot p_W \right)\left(p_b \cdot p_Z \right) \right]\nonumber\\
&&+{4 \over m_Z^2}\left( 1 - \gamma \right)
\left(k_2 \cdot p_Z\right)\left[ 2 \left(p_b \cdot p_Z
\right) \left(p_t \cdot p_Z \right) + \left(p_b \cdot p_Z \right)\left(p_t
\cdot p_W \right) 
\right. \nonumber\\ 
&&\left.\qquad\qquad\qquad\qquad\qquad\qquad
- \left(p_W \cdot p_Z \right)\left(p_t \cdot p_b \right)
+ \left(p_t \cdot p_Z \right)\left(p_b \cdot p_W \right) \right]\nonumber\\
&&-{{2} \over m_W^2}\left( 3 + \gamma \right)
\left(p_W \cdot p_Z \right)\left[
\left(k_2 \cdot p_W\right)\left(p_t \cdot p_b \right) - \left(p_b \cdot p_W
\right)\left(k_2 \cdot p_t \right) + \left( p_t \cdot p_W \right) \left(k_2
\cdot p_b \right) \right] \nonumber\\
&&+{{2} \over m_Z^2}\left( 3 - \gamma \right)
\left(p_W \cdot p_Z \right)\left[
\left(k_2 \cdot p_Z\right)\left(p_t \cdot p_b \right) - \left(p_t \cdot p_Z
\right)\left(k_2 \cdot p_b \right) + \left( p_b \cdot p_Z \right) \left(k_2
\cdot p_t \right) \right] \nonumber\\
&&+{4 \over {m_W^2 m_Z^2}}\left(1 + \gamma \right)
\left(p_W \cdot p_Z\right)\left(p_t \cdot p_W \right)
\left[\left(k_2 \cdot p_W \right)\left(p_b \cdot p_Z \right) - \left(p_W \cdot
p_Z \right) \left(k_2 \cdot p_b \right) + \left(p_b \cdot p_W \right) \left(k_2
\cdot p_Z \right) \right]\nonumber\\
&&-{4 \over {m_W^2 m_Z^2}}\left(1 - \gamma \right)
\left(p_W \cdot p_Z\right)\left(p_b \cdot p_Z \right)
\left[\left(k_2 \cdot p_W \right)\left(p_t \cdot p_Z \right) - \left(p_W \cdot
p_Z \right) \left(k_2 \cdot p_t \right) + \left(p_t \cdot p_W \right) \left(k_2
\cdot p_Z \right) \right]\bigg\}\nonumber\\
&&+4m_b^2 g_{b_R} 
\left\{ \left(-4 + \gamma \right)
\left(p_t \cdot p_W \right) + \left(-2 + \gamma \right) \left(p_t \cdot p_Z
\right) 
+ {{1} \over {m_W^2}}\left( {7 \over 2} + {1 \over 2} \gamma \right)
\left(p_W \cdot p_Z \right)\left(p_t \cdot p_W
\right) \right.\nonumber\\
&&\left.\qquad\qquad
+{1 \over m_Z^2}\left( {3 \over 2} - {1 \over 2} \gamma \right) 
\left(p_W \cdot p_Z \right)\left(p_t \cdot p_Z \right)
+{1 \over {m_W^2 m_Z^2}}\left( 1 + \gamma \right)
\left(p_W \cdot p_Z \right)^2 \left(p_t \cdot p_W
\right)\right\}\ ,
\end{eqnarray}
where $\gamma = 1 - m_Z^2/m_W^2$ or $\sin^2\theta_W$.
The interference terms
${\cal A}_2 {\cal A}_3^*$ and ${\cal A}_1 {\cal A}_3^*$ are related
by the interchanges $g_{t_{L,R}} \leftrightarrow g_{b_{L,R}}$,
$m_t \leftrightarrow m_b$, $p_t \leftrightarrow p_b$, and 
$k_1 \leftrightarrow k_2$.

The above square amplitudes have been written in terms of $k_1$ and $k_2$,
and $p_t$ and $p_b$,
in order to exhibit the symmetries of the square amplitudes explicitly.
The total square amplitude can be rewritten in terms of the three dot products
$(p_b \cdot p_W)$, $(p_b \cdot p_Z)$, and $(p_W \cdot p_Z)$, by eliminating
$p_t$, $k_1$ and $k_2$ in the above formulae.

The partial width for the decay mode $t \rightarrow b W^+ Z$ is given by
the three-body phase space integral
\begin{equation}
\Gamma\left(t \rightarrow b W^+ Z \right) = {1 \over {(2 \pi)}^3}
{1 \over {32 m_t^3}}\  \int dm_{23}^2\ dm_{12}^2 \ \overline{|{\cal A}|^2}
\end{equation}
where the invariant square masses $m_{ij}^2 = (p_i + p_j)^2$ are defined
in terms of the momenta of the final particles, and the
spin-averaged square amplitude
\begin{equation}
\overline{|{\cal A}|^2} = \frac 1 2 \ |{\cal A}|^2,
\end{equation}
since one averages rather than sums over the top quark spin.

The partial decay width $\Gamma\left( t \rightarrow b W^+ Z \right)$ 
is plotted in Fig.~2 as a function of the top quark mass.  
The phase
space integral was performed numerically for the parameter values
$m_W = 80.3$~GeV, $m_Z =91.2$~GeV, $m_b = 4.5$~GeV, $\sin^2 \theta_W = 0.23$
and $|V_{tb}|=1$.  
The partial width 
is plotted over the range from
$m_t = 176$~GeV, where the partial width vanishes, to $m_t = 200$~GeV,
where the partial decay width is $2.5 \times 10^{-5}$~GeV.  
The branching ratio ${\rm BR}(t \rightarrow b W^+ Z)$ 
also is plotted as a function
of the top quark mass in Fig.~3, assuming that 
the total
width of the top quark is dominated by $t \rightarrow b W^+$,  
\begin{eqnarray}
\Gamma\left(t \rightarrow b W^+\right) &&= |V_{tb}|^2\  {g^2 \over {64 \pi}}
{1 \over {m_W^2 m_t^3}} \ \lambda^{1 \over 2}\left(m_t^2,m_W^2,m_b^2\right)
\nonumber\\
&&\qquad\left\{m_t^4 + m_b^4 -2 m_W^4 + m_t^2 m_W^2 + m_b^2 m_W^2 
-2 m_t^2 m_b^2 \right\},
\end{eqnarray}
where
\begin{equation}
\lambda( x,y,z) = x^2 + y^2 + z^2 -2xy -2yz -2xz.
\end{equation}
 The CKM matrix element
$|V_{tb}|^2$ cancels out of the branching ratio.  The branching ratio 
increases from zero for $m_t = 176$~GeV
to $1.0 \times 10^{-5}$ for $m_t = 200$~GeV.\footnote{The value of 
the branching ratio for this value of $m_t$ 
is consistent with the result of Mahlon and Parke\cite{mahlon}
in the narrow width approximation.  There 
is some numerical difference with their narrow width
results for smaller values
of $m_t$, which probably stems from the inclusion of finite width
effects proportional to $\Gamma_W/m_W$ and $\Gamma_Z/m_Z$
in their calculation.}  
Although this branching
ratio is too small to be observed at the Tevatron, it is large enough
to be interesting
for the LHC which is expected to yield about a million fully-reconstructed top
quark events per year\cite{atlas}.  The observability of this decay mode
depends on the precise value of the top quark mass.  The branching ratio
is greater than $10^{-6}$ for $m_t 
\ \raise.3ex\hbox{$>$\kern-.75em\lower1ex\hbox{$\sim$}}\ 
187$~GeV.  The extreme sensitivity of the branching ratio to the top quark
mass implies that the decay mode could be used to extract or bound 
the top quark mass.  The decay mode also is sensitive to the presence of
the triple gauge vertex $W^+ W^- Z$ with the standard model coupling.

The rare decay $t \rightarrow b W^+ Z$ is at threshold for the present
central value of the top quark mass, so the decays $t \rightarrow s W^+ Z$
and $t \rightarrow d W^+ Z$ could be more important if
$t \rightarrow b W^+ Z$ is kinematically forbidden or just allowed.  Alternatively,
it might be possible to look at these modes by applying a tight cut on the 
invariant mass of the $W^+$ and $Z$ momenta to exclude $t \rightarrow b W^+ Z$
but not $t \rightarrow s W^+ Z$ and $d W^+ Z$.  
The partial decay widths for the $s$ and
$d$ final states can be obtained from the partial decay width for the $b$ mode
by replacing $m_b$ by $m_s$ or $m_d$, and $|V_{tb}|^2$ by $|V_{ts}|^2$ 
or $|V_{td}|^2$.  The partial decay width 
$\Gamma\left( t \rightarrow d_i W^+ Z \right)/ |V_{ti}|^2$ is plotted as
a function of the top quark mass for vanishing $m_{d_i}$ in Fig.~4.  The
partial width divided by the CKM matrix element squared is zero at threshold
where $m_t = m_W + m_Z$ and increases to $2.7 \times 10^{-5}$~GeV at 
$m_t=200$~GeV.  For canonical values of $|V_{ts}|^2$ and $|V_{td}|^2$,
these partial widths will be too small to be observed at LHC. 

\section{$t \rightarrow c W^+ W^-$}

The rare decay $t \rightarrow c W^+ W^-$ proceeds through
tree-level graphs with intermediate $d$, $s$, and $b$ quarks,
as depicted in Fig.~5.  The amplitude for the decay is
\begin{equation}
{\cal A} = \sum_{j=d,s,b} V_{tj} V_{cj}^*  
\left({{ig} \over \sqrt{2}}\right)^2
\epsilon_{W^-}^\mu \epsilon_{W^+}^\nu \ 
\bar u(p_c) \left[ \gamma^\mu P_L \left({i \over {{\rlap{$k$}/} - m_j}}\right)
\gamma^\nu P_L \right] 
u(p_t), \\
\end{equation}
where
$k= p_c + p_{W^-}= p_t - p_{W^+}$.  The spin-averaged amplitude squared
is given by
\begin{eqnarray}
\overline{ |{\cal A}|}^2 =&& 
{ 1 \over 2} \sum_{j,k} V_{tj} V_{cj}^* V_{tk}^* V_{ck}
\left( {g \over \sqrt{2} } \right)^4
\left({1 \over {k^2 - m_j^2}} \right)
\left({1 \over {k^2 - m_k^2}} \right) \times \nonumber\\
4 \ \Bigg\{ &&\left[ \left( k \cdot p_t \right) \left( k \cdot p_c \right)
-{1 \over 2} k^2 \left( p_c \cdot p_t \right) \right] \nonumber\\ 
&&+ {2 \over {m_W^2}} \bigg( \left( p_t \cdot p_{W^+} \right)
\left[ \left( k \cdot p_c \right) \left( k \cdot p_{W^+} \right) 
- {1 \over 2} k^2 \left( p_c \cdot p_{W^+} \right)
\right] \nonumber\\
&&\qquad\quad + \left( p_c \cdot p_{W^-} \right) 
\left[ \left(k \cdot p_t \right) \left( k \cdot p_{W^-} \right)
 - {1 \over 2} k^2 \left( p_t \cdot p_{W^-} \right) \right] \bigg) \\
&&+ {4 \over m_W^4 } \left( p_t \cdot p_{W^+} \right)
\left( p_c \cdot p_{W^-} \right) 
\left[ \left( k \cdot p_{W^+} \right) 
\left( k \cdot p_{W^-} \right) - {1 \over 2} k^2 
\left( p_{W^+} \cdot p_{W^-} \right) \right]
\Bigg\}, \nonumber
\end{eqnarray}
where the factor of $1/2$ comes from averaging over the top quark spin.  
Note that this square amplitude can be derived from Eq.~(8) or~(9).
The 
amplitude squared can be rewritten in terms of the three dot products
$(p_c \cdot p_{W^+} )$, $(p_c \cdot p_{W^-})$ and $(p_{W^+} \cdot p_{W^-})$
of the final particle momenta
by eliminating $p_t$ and $k$ in the above formula.

The partial width for the decay mode $t \rightarrow c W^+ W^-$
is given by the three-body phase space integral
\begin{eqnarray}
\Gamma\left( t \rightarrow c W^+ W^- \right) &=& 
{1 \over {\left(2\pi\right)^3}} \ {1 \over {32 m_t^3}} 
\int dm_{23}^2 \ dm_{12}^2 \ \overline{ |{\cal A}| }^2,
\end{eqnarray}
where the invariant square masses
$m_{ij}^2 = (p_i + p_j)^2$ are defined in terms of the momenta of
the final particles.  It is possible to do the
$m_{W^+W^-}^2$ integral explicitly, so that the partial width is
given by an integral over
$x= m_{cW^-}^2 = (p_c + p_{W^-})^2 = (p_t - p_{W^+})^2 = k^2$,
\begin{eqnarray}
\Gamma\left( t \rightarrow c W^+ W^- \right) &&= 
{1 \over {\left(2\pi\right)^3}}\ {1 \over {32 m_t^3}}
\ { 1 \over 2} \ \left( {g \over \sqrt{2} } \right)^4
\sum_{j,k} V_{tj} V_{cj}^* V_{tk}^* V_{ck} \times \nonumber\\
\int_{(m_c + m_W )^2}^{(m_t - m_W)^2} dx && 
\left( {1 \over {x - m_j^2} } \right) \left( {1 \over {x - m_k^2} } \right)
{1 \over {2x}} \lambda^{{1 \over 2}}\left( x, m_c^2, m_W^2 \right)
\lambda^{{1 \over 2}}\left( x, m_t^2, m_W^2 \right) \times
\nonumber\\
&&\Bigg\{ \left( x + m_t^2 - m_W^2 \right) \left( x + m_c^2 -m_W^2 \right)
\\
&&+ {1 \over m_W^2} \left[ \left(x -m_t^2 -m_W^2 \right) 
\left( x + m_c^2 - m_W^2 \right)\left(x - m_t^2 + m_W^2 \right)
\right. \nonumber\\
&&\qquad\quad
+ \left. \left(x -m_c^2 - m_W^2 \right) \left(x + m_t^2 - m_W^2 \right)
\left( x - m_c^2 + m_W^2 \right) \right]\nonumber\\
&&+ { 1 \over m_W^4} \left[ \left(x - m_t^2 + m_W^2 \right)
\left(x - m_t^2 - m_W^2 \right) \left( x -m_c^2 + m_W^2 \right) 
\left( x - m_c^2 -m_W^2 \right) \right] \Bigg\}. \nonumber
\end{eqnarray}
The integrand of Eq.~(21) is symmetric under the interchange
$m_t^2 \leftrightarrow m_c^2$, but this symmetry is
broken by the  
limits of integration of the remaining phase space integral.  
An important observation
about the decay width is that the
width vanishes for $m_j^2=0$ or $m_k^2=0$ by CKM unitarity, 
\begin{equation}
\sum_{j=d,s,b} V_{tj} V_{cj}^* =0 \ .
\end{equation}
This GIM suppression can be made manifest by replacing
the two $d$-quark propagators by
\begin{equation}
\left( { 1 \over {x - m_j^2} } \right) \rightarrow 
\left[ \left( { 1 \over {x - m_j^2} } \right) - {1 \over x} \right]
= {m_j^2 \over {x \left(x - m_j^2 \right) }}, 
\end{equation}
which implies that the integrand is multiplied by 
\begin{equation}
{{m_j^2 m_k^2} \over x^2}\ .
\end{equation}
Thus, the final formula for the partial width is
\begin{equation}
\Gamma\left( t \rightarrow c W^+ W^- \right) = {1 \over {\left(2\pi\right)^3}}
\ {1 \over {32 m_t^3}}\ 
{ 1 \over 2} \ \left( {g \over \sqrt{2} } \right)^4 \ 
\sum_{j,k} V_{tj} V_{cj}^* V_{tk}^* V_{ck}
\ I\left( m_j^2, m_k^2, m_c^2, m_t^2, m_W^2 \right),
\end{equation}
where the integral equals
\begin{eqnarray}
&&I\left( m_j^2, m_k^2, m_c^2, m_t^2, m_W^2 \right) = m_j^2 m_k^2\  
\int_{(m_c + m_W )^2}^{(m_t - m_W)^2} dx \ 
\left( {1 \over {x - m_j^2} } \right) \left( {1 \over {x - m_k^2} } \right)
\times \nonumber\\
&&\qquad
{1 \over {2x^3}} \lambda^{{1 \over 2}}\left( x, m_c^2, m_W^2 \right)
\lambda^{{1 \over 2}}\left( x, m_t^2, m_W^2 \right)
\Bigg\{ \left( x + m_t^2 - m_W^2 \right) \left( x + m_c^2 -m_W^2 \right)
\nonumber\\
&&\qquad
+ {1 \over m_W^2} \left[ \left(x -m_t^2 -m_W^2 \right) 
\left( x + m_c^2 - m_W^2 \right)\left(x - m_t^2 + m_W^2 \right)
\right. \\
&&\qquad
\qquad\quad+ \left. 
\left(x -m_c^2 - m_W^2 \right) \left(x + m_t^2 - m_W^2 \right)
\left( x - m_c^2 + m_W^2 \right) \right]\nonumber\\
&&\qquad
+ { 1 \over m_W^4} \left[ \left(x - m_t^2 + m_W^2 \right)
\left(x - m_t^2 - m_W^2 \right) \left( x -m_c^2 + m_W^2 \right) 
\left( x - m_c^2 -m_W^2 \right) \right] \Bigg\}. \nonumber
\end{eqnarray}

Numerical integration of Eq.~(26) (which assumes that GIM suppression
is operative)
shows that the decay width is completely dominated by the contribution 
with $m_j^2 = m_k^2 =m_b^2$.  
The partial width is plotted as a function
of the top quark mass  
for $m_W=80.3$~GeV, $m_c=1.5$~GeV,
$\sin^2 \theta_W = 0.23$ and $V_{cb} = 0.036-0.046$.  The two curves 
correspond to the lower and upper values of the CKM matrix element $V_{cb}$.
The partial width vanishes at threshold where $m_t = m_c + 2 m_W$, and 
is at most $\approx 10^{-12}$~GeV for $m_t=200$~GeV.  
This extremely small partial width is a direct
consequence of three-family unitarity of the CKM matrix in the $u$-quark
sector.  If the GIM
suppression condition Eq.~(22) is relaxed, the integral appearing
in Eq.~(21) is a factor of $2 \times 10^5$ larger than 
$I(m_b^2,m_b^2,m_c^2,m_t^2,m_W^2)$ for each value of $m_j^2$ and $m_k^2$.
Thus, it is quite possible that the rare decay $t \rightarrow c W^+ W^-$
occurs at an observable level in non-standard model theories.  A search
for this rare decay mode would directly test CKM unitarity
of the $tc$ rows of the CKM matrix.    

The partial width for the rare decay $t \rightarrow u W^+ W^-$ can be
obtained from the above with the replacement $c \leftrightarrow u$.
The partial width for the up mode is even smaller than for
the charm mode due to smaller CKM matrix elements.  This decay mode
can be used to test CKM unitarity of the $tu$ rows of the CKM matrix.

\section{Conclusions}

The partial widths for the rare top decay modes $t \rightarrow b W^+ Z$,
$s W^+ Z$, $d W^+ Z$, $c W^+ W^-$ and $u W^+ W^-$ have been calculated in
the standard model.  The decay mode $t \rightarrow b W^+ Z$ is potentially
observable at LHC rates for top quark masses above $187$~GeV, and could
be used to accurately determine the top quark mass.  The decay amplitude
also depends on the triple decay vertex $W^+ W^- Z$, and therefore tests 
for the
presence of this coupling and its value.  
The decays
$t \rightarrow c W^+ W^-$ and $u W^+ W^-$ are extremely GIM-suppressed in
the standard model, but may be much larger in non-standard scenarios.
A search for these rare decay modes tests CKM unitarity of the $tc$ and
$tu$ rows of the CKM matrix,
\begin{equation}
\sum_{j=d,s,b} V_{tj} V_{u_i j}^* =0 \ .
\end{equation}

\section*{Acknowledgments}

This work was supported in part by the Department of Energy
under grant DOE-FG03-90ER40546.  E.J. also was supported in part by NYI
award PHY-9457911 from the National Science Foundation and by a research
fellowship from the Alfred P. Sloan Foundation.

\vfill\break\eject

\begin{figure}
\caption{$t \rightarrow b W^+ Z$.  Feynman diagrams correspond to
the amplitudes ${\cal A}_1$, ${\cal A}_2$, and ${\cal A}_3$.}
\label{fig:feynmandiagrams}
\end{figure}

\begin{figure}
\caption{$\Gamma(t \rightarrow b W^+ Z)$ as a function of the top quark mass
for $m_W = 80.3$~GeV, $m_Z= 91.2$~GeV, $m_b=4.5$~GeV, 
$\sin^2 \theta_W =0.23$, and $|V_{tb}|^2 = 1$.  The partial decay width
vanishes at threshold where $m_t = m_b + m_W + m_Z$.}
\label{fig:gamma}
\end{figure}

\begin{figure}
\caption{$BR(t \rightarrow b W^+ Z)$ as a function of the top quark mass
for $m_W = 80.3$~GeV, $m_Z= 91.2$~GeV, $m_b=4.5$~GeV, and 
$\sin^2 \theta_W =0.23$.  The branching ratio
vanishes at threshold where $m_t = m_b + m_W + m_Z$.}
\label{fig:br}
\end{figure}

\begin{figure}
\caption{$\Gamma(t \rightarrow d_i W^+ Z)/ |V_{ti}|^2$ as a function 
of the top quark mass
for $m_W = 80.3$~GeV, $m_Z= 91.2$~GeV, 
$\sin^2 \theta_W =0.23$ and $m_{d_i}=0$.  The partial decay width
vanishes at threshold where $m_t = m_W + m_Z$.  This graph is relevant
for the decays $t \rightarrow d W^+ Z$ and $t \rightarrow s W^+ Z$.}
\label{fig:gammazero}
\end{figure}

\begin{figure}
\caption{$t \rightarrow c W^+ W^-$.}
\label{fig:feyndiag}
\end{figure}

\begin{figure}
\caption{$\Gamma(t \rightarrow c W^+ W^-)$ as a function of the top quark mass
for $m_W = 80.3$~GeV, $m_c=1.5$~GeV, 
$\sin^2 \theta_W =0.23$ and $V_{cb}=0.036-0.046$.  The partial decay width
vanishes at threshold where $m_t = m_c + 2 m_W$.}
\end{figure}

\vfill\break\eject

\centerline{Figure 1}
\vskip1.truein

\epsfxsize=8cm
\epsffile{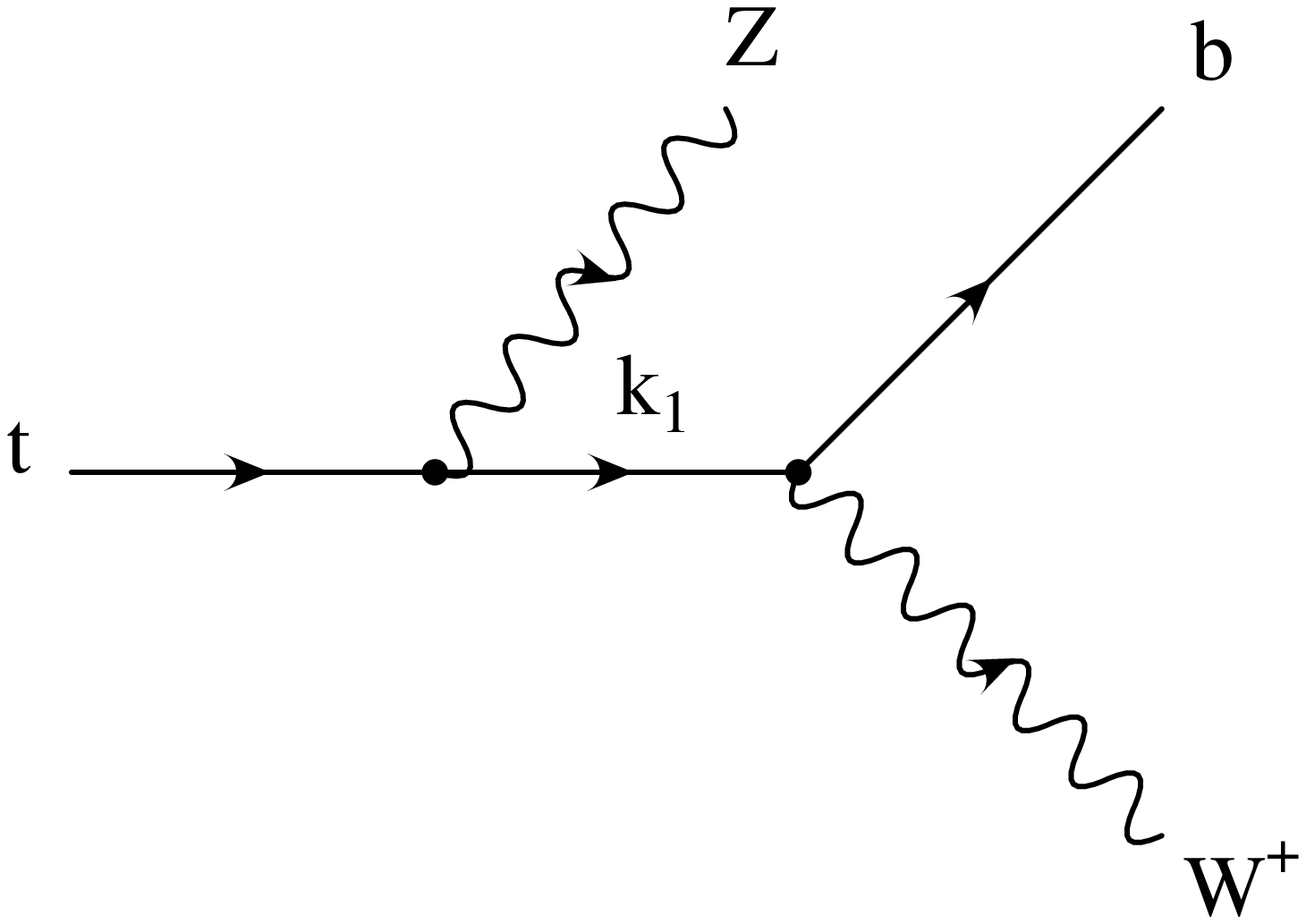}
\medskip

\epsfxsize=7.0cm
\epsffile{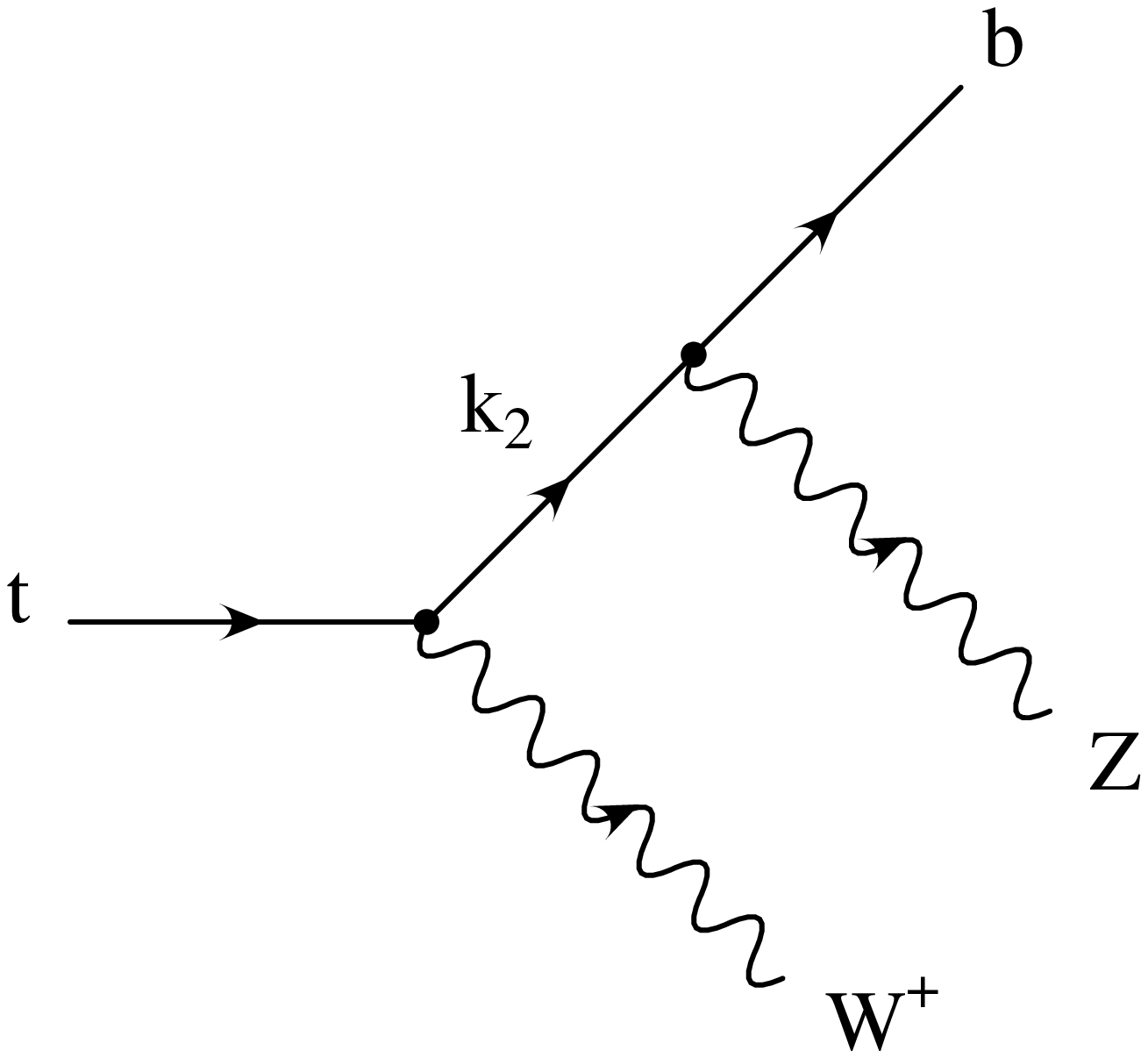}
\medskip

\epsfxsize=8cm
\epsffile{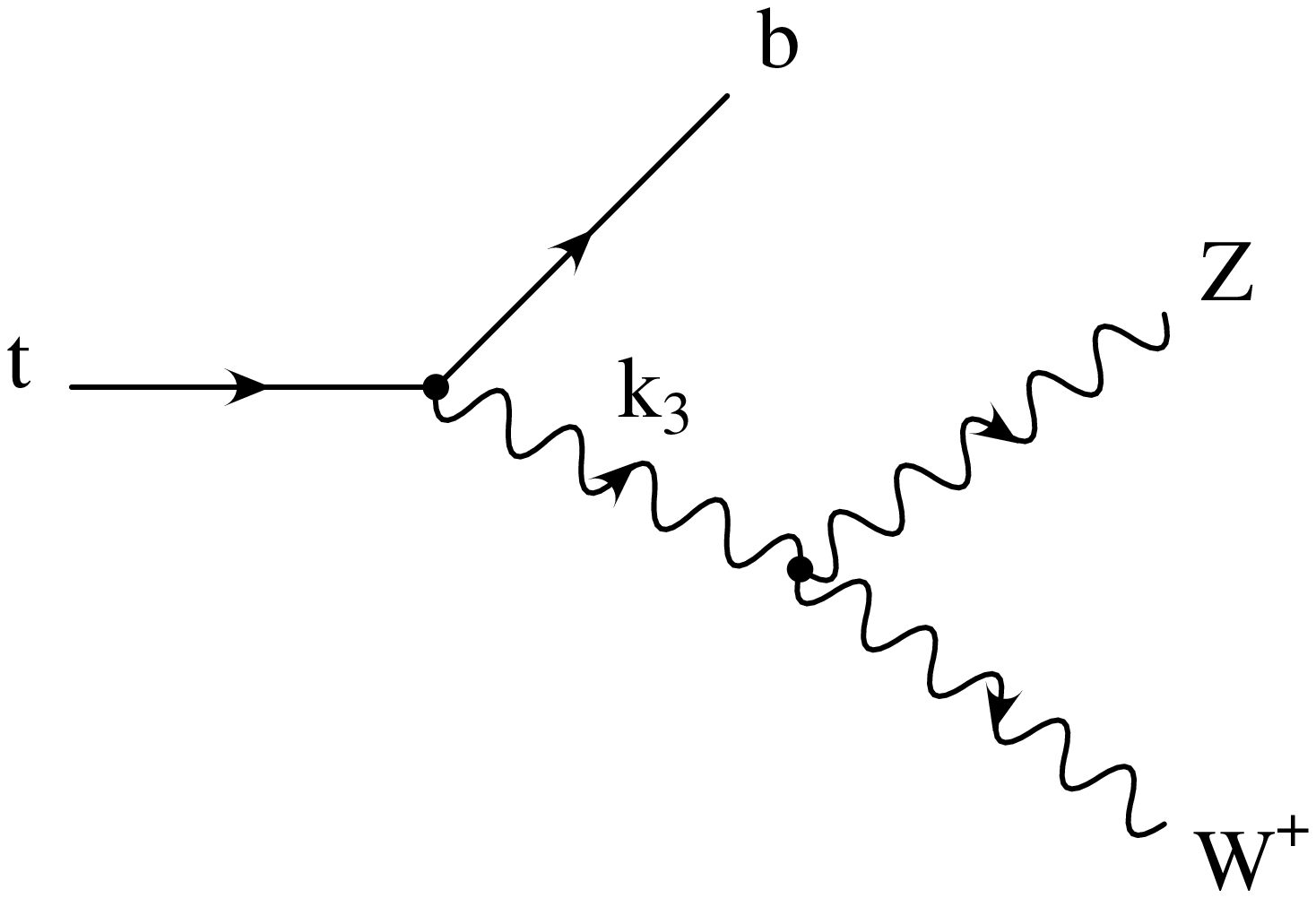}
\medskip

\vfill\break\eject

\epsffile{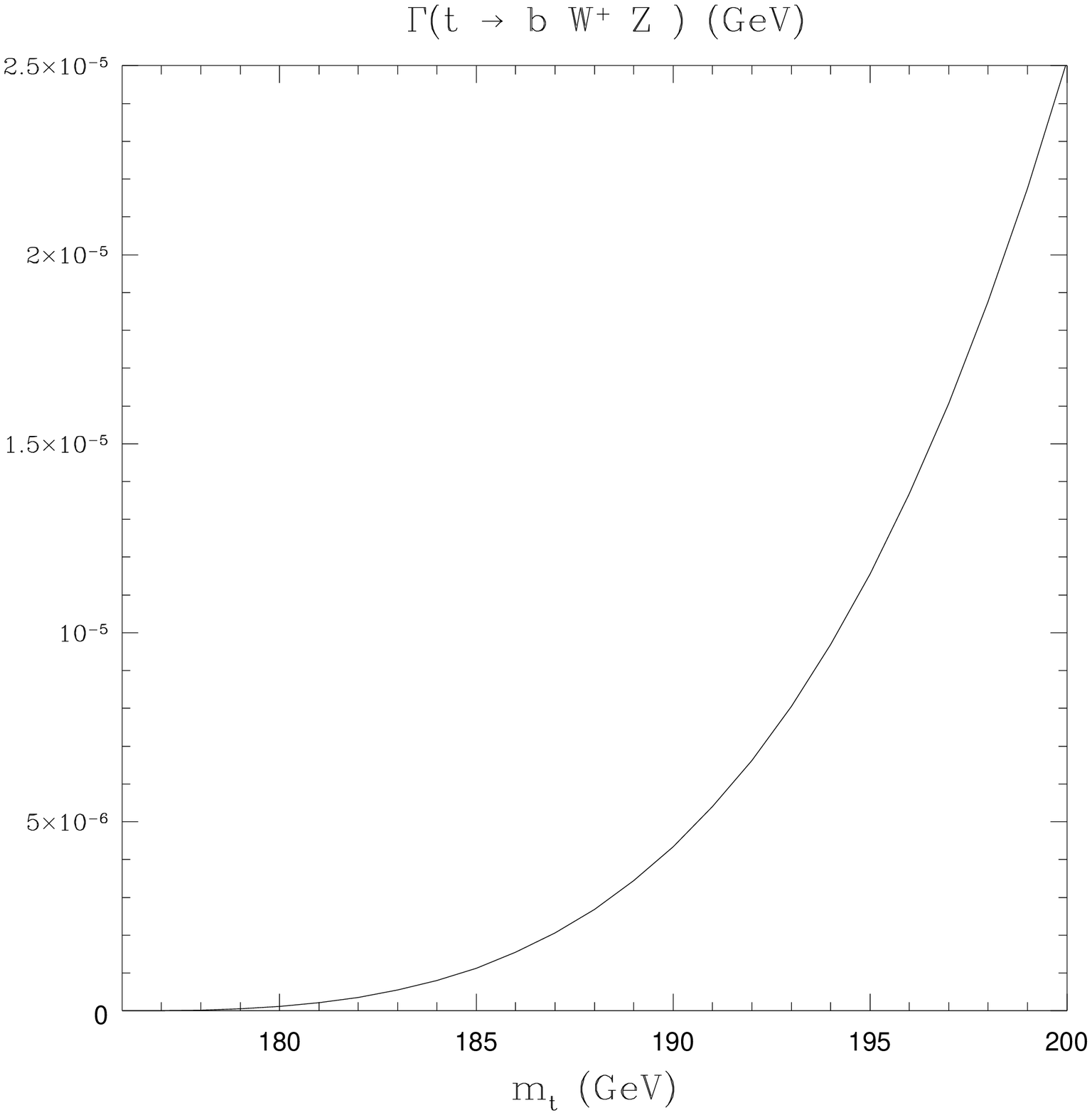}

\epsffile{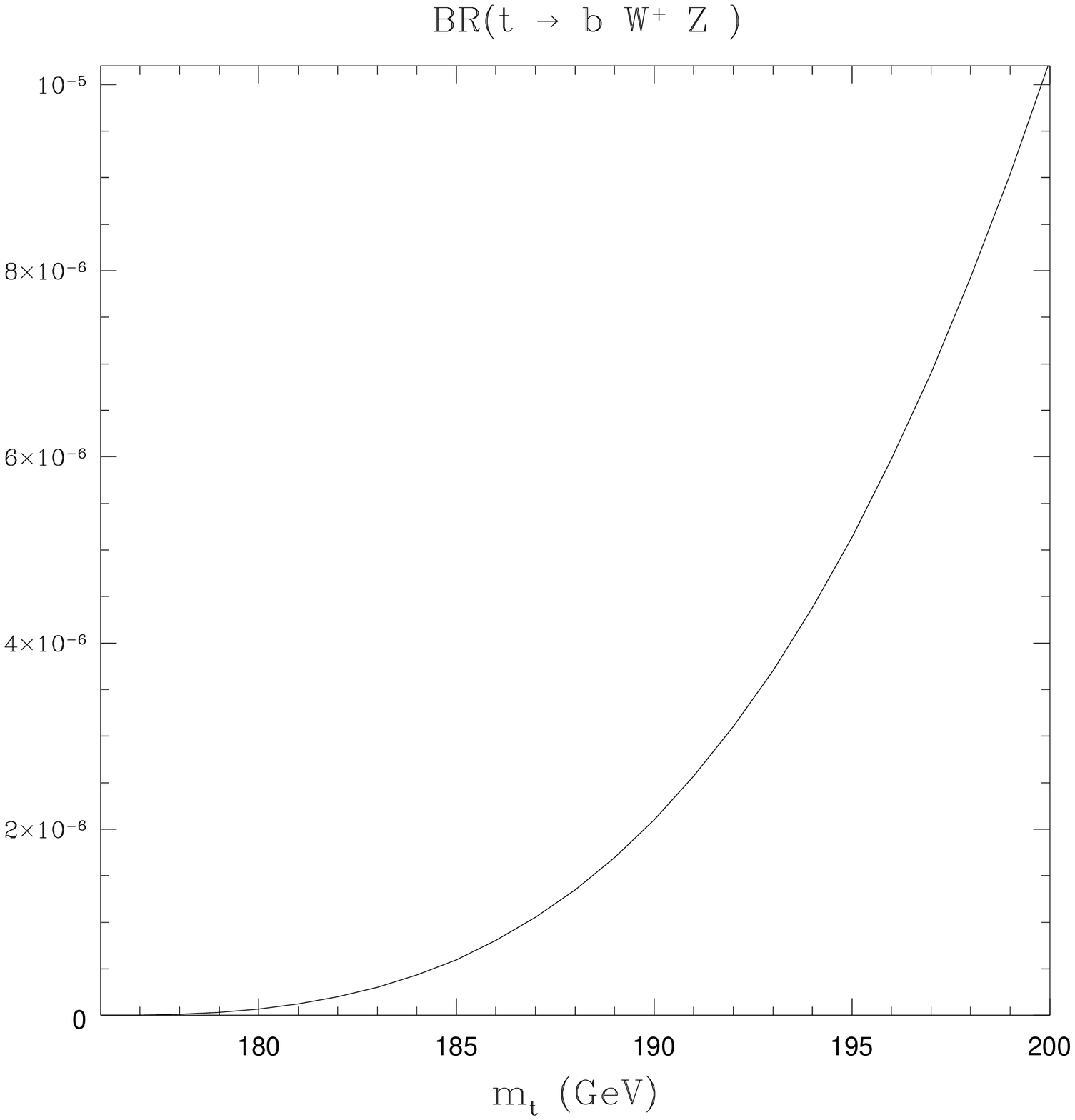}

\epsffile{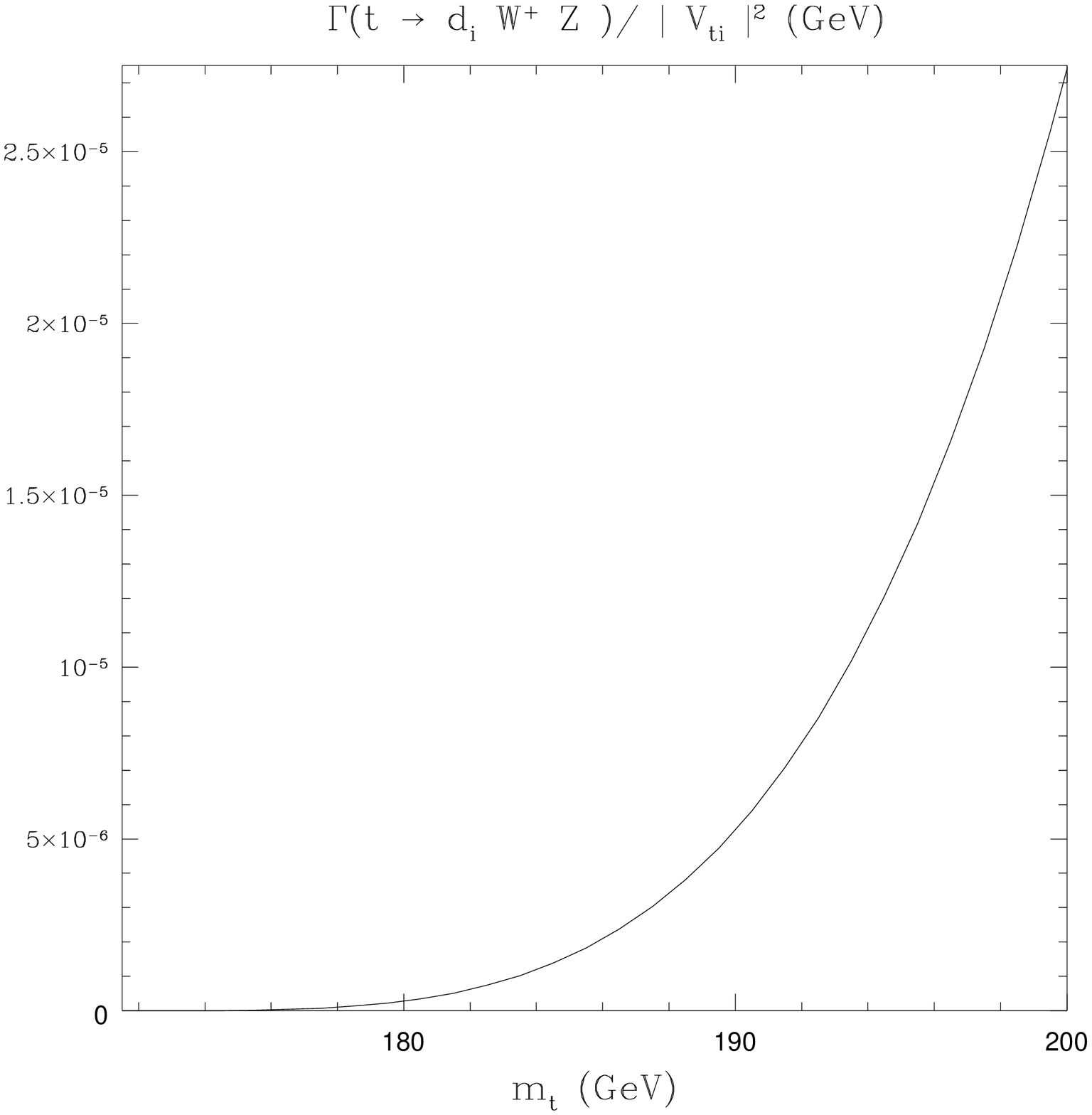}

\centerline{Figure 5}
\vskip1.truein

\epsfxsize=8cm
\epsffile{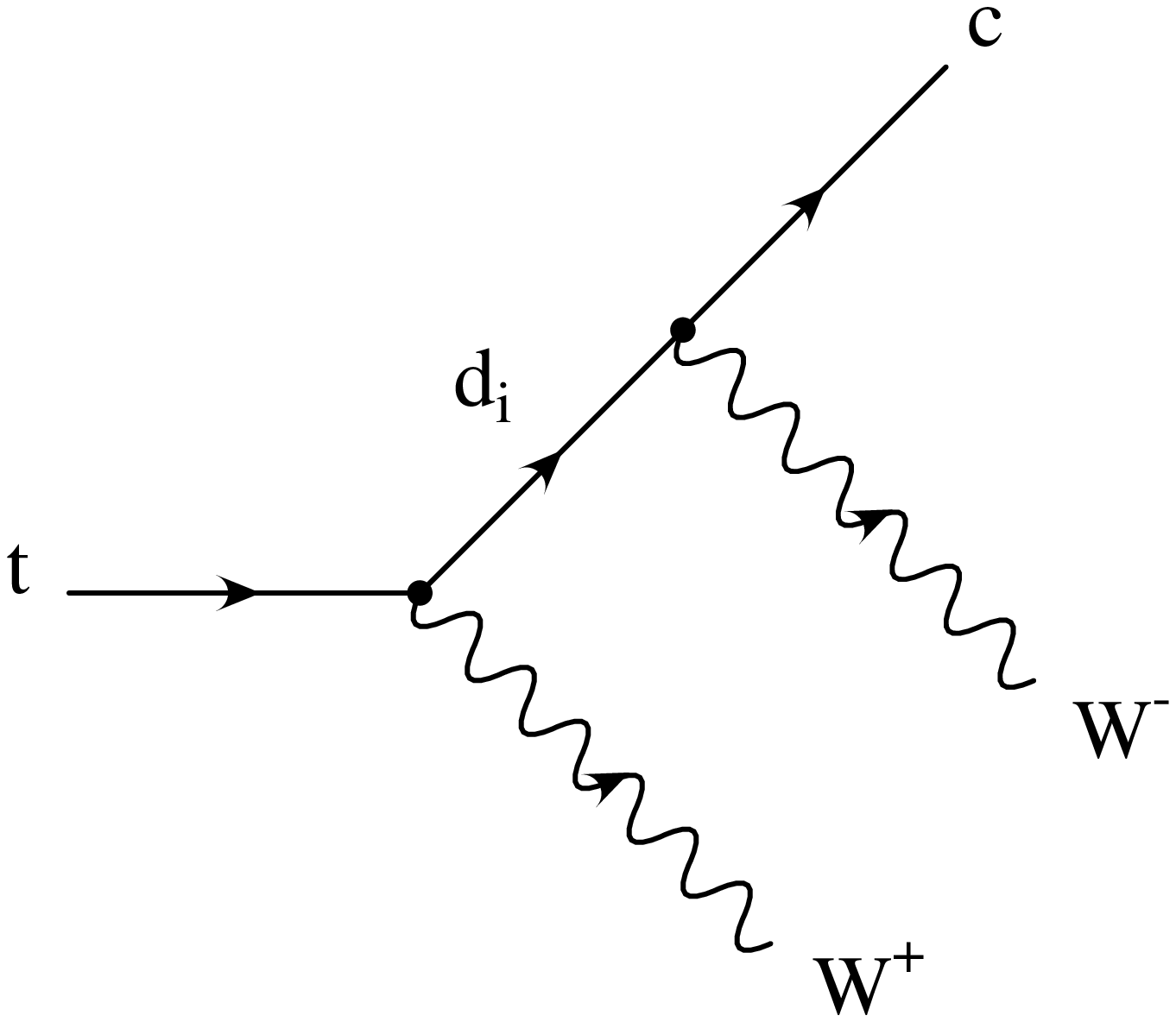}
\medskip

\vfill\break\eject

\epsffile{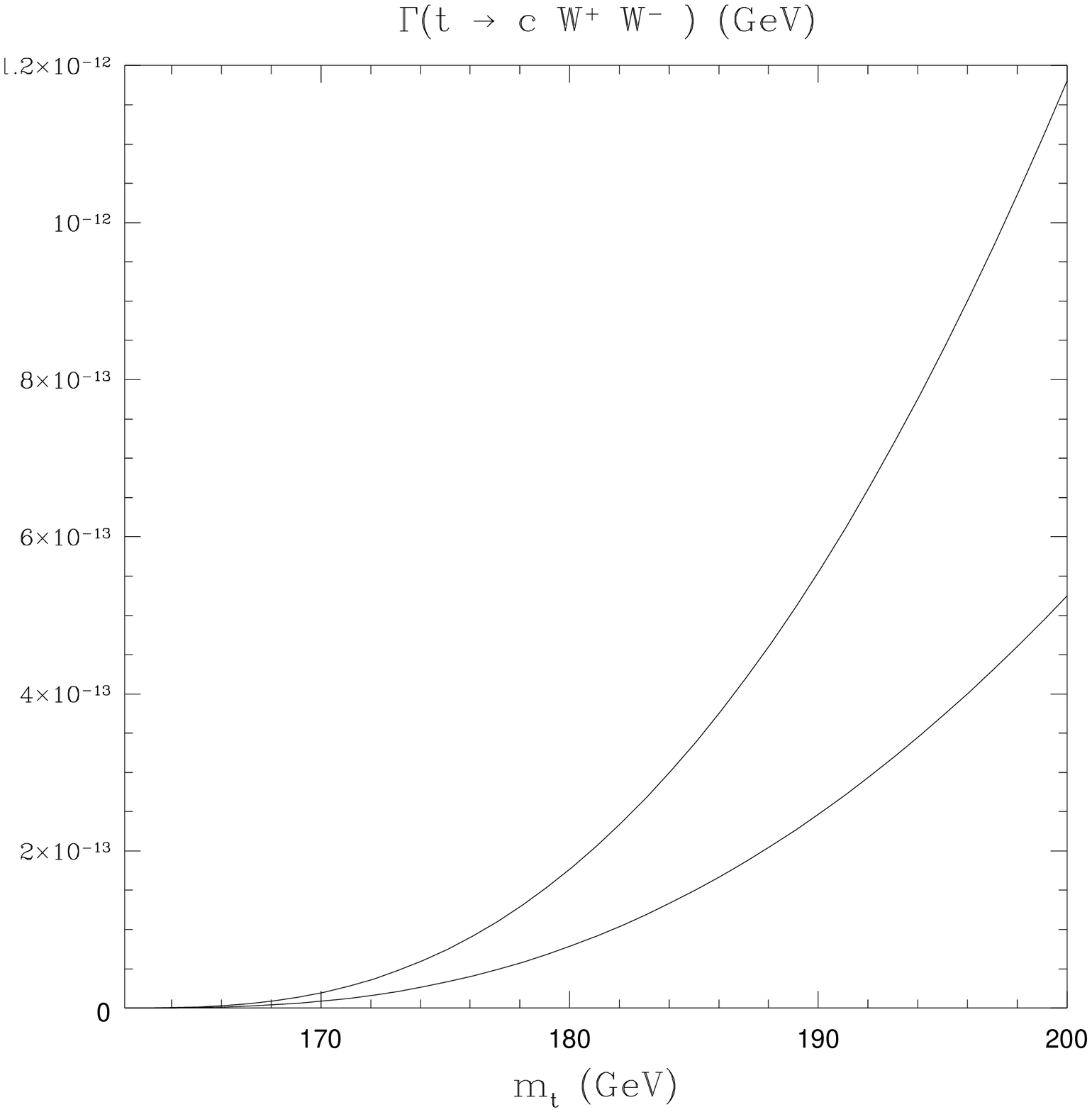}

\end{document}